\documentclass{rsproca}

\usepackage{epsfig, amsmath, amsfonts, graphicx, amssymb}

\newcommand\bb[1] {   \mbox{\boldmath{$#1$}}  }

\newcommand\del{\bb{\nabla}}

\newcommand\beq{ \begin{equation} }
\newcommand\eeq{ \end{equation} }

\begin{document}

\title{Dynamical, biological, and anthropic consequences of equal lunar and solar angular radii}

\author{Steven A. Balbus}    

\address{$^{1}$Astrophysics, Keble Road, Denys Wilkinson Building,
University of Oxford, Oxford OX1 3RH}  

\subject{Astronomy, Astrobiology}

\keywords{Tides, evolution, Moon} 

\corres{Steven Balbus\\ 
\email{steven.balbus@astro.ox.ac.uk}}

\begin{abstract}

The nearly equal lunar and solar angular sizes as subtended at the Earth 
is generally regarded as a coincidence.   This is, however, an incidental
consequence of the tidal forces from these bodies being comparable.  
Comparable magnitudes implies strong temporal modulation, as the forcing 
frequencies are nearly but not precisely equal.
We suggest that on the basis of paleogeographic reconstructions,
in the Devonian period, when the first tetrapods appeared on land,
a large tidal range would accompany these modulated tides.   This would have been conducive
to the formation of a network of isolated tidal pools, lending support to 
A.S. Romer's classic idea that the evaporation of shallow pools was an evolutionary 
impetus for the development of chiridian limbs in aquatic tetrapodomorphs.
Romer saw this as the reason for the existence of limbs,
but strong selection pressure for terrestrial navigation would
have been present even if the limbs were aquatic in origin.
Since even a modest difference in the
Moon's angular size relative to the Sun's 
would lead to a qualitatively different tidal modulation,
the fact that we live on a planet with a Sun and Moon of close apparent size
is not entirely coincidental: it
may have an anthropic basis.  

\end{abstract}


\begin{fmtext}

\end{fmtext}
\maketitle

\section {Introduction}

The fact that the solid angles subtended by the Sun and Moon at
the Earth are nearly equal is generally regarded as a coincidence, one that leads
to the phenomenon of total solar eclipses of
agreeably short duration.  The purpose of ths article is to
propose an anthropic explanation for why the near equality of solar
and lunar angular sizes may not be entirely coincidental, and in the
process lend support to an old idea of A. S. Romer [1] on the evolutionary role
of tidal pools.  The idea is that
the near match of angular sizes is a
mathematical, but incidental, by-product of the presence of a strongly
modulated tidal forcing by the Sun and Moon.  It is the latter
that is the true biological imperative.

The two earliest known tetrapods with more than fragmentary remains, {\it Acanthostega}
and {\it Ichthyostega}, are thought to have been fully (perhaps only
predominantly in the case of {\em Ichthyostega}) aquatic creatures [2].
The coastal and estuarial waters such organisms and their immediate ancestors are believed to have
inhabited would have been subject to sizeable and irregular tides, leaving an
inland network of pools.  The farthest
inland of these pools would on occasion have been left exposed weeks at time,
utimately evaporating.  A creature caught in one
of these isolated inland pools would consequently have faced 
dehydration or suffocation.
But given a sense of direction to its flailing, there would have been
plenty of inviting pools closer to the sea.  These would be refreshed
progessively more often than pools deeper inland.  In addition, any
fish left stranded in these inland pools
would have been easy prey for those
predators adapted to directed terrestrial motion [2].
The exigencies and advantages of large motor control in
a network of tidal pools would surely have been an important
evolutionary impetus to evolve weight-bearing chiridian limbs.
This simple and important argument, which
ties together evolutionary biology with established
tidal theory, deserves to be much more widely known. 
In this paper, we illustrate by explicit calculation
the sensitivity of the equilibrium tidal height to the relative
lunar and solar angular sizes, and
point out that the continental configuration associated with Devonian plate tectonics may have
been particularly conducive to a large tidal range.

The argument associating tidal modulation with perceived lunar and solar angular sizes
is very simple.  Isaac Newton himself was aware of it: in the {\it Principia,}
he shows that one may use the ratio of the spring to neap tides to deduce 
the fact that the Moon's
tidal force is stronger than the Sun's, and that the Moon therefore must be the
denser body.  The relation is also clearly discussed
in an unpublished 1992 manuscript by C. Tolbert and C.
Sarazin [3], which goes on to note some of the biological implications that
we discuss more fully in this work.

The tidal force, which arises from the quadrupole term of the two-body
potential of the disturbing source and its host, is proportional to the mass
of the disturber and the reciprocal cube of the distance between the
centres-of-mass.  Assume for the moment that the Sun exerts a tidal
force on the Earth which is a fraction $f$ of the Moon's tidal force.   Then,
\beq
{M_S\over r_S^3} = f{M_m\over r_m^3}
\eeq
where $M_S$ is the mass of the Sun, $r_S$ the Earth-to-Sun centre-of-mass distance,
and $M_m$ and $r_m$ the corresponding quantities for the Moon.  The masses, in turn,
are proportional to the average internal density times the cube of the body's diameter.
With $\rho$ and $D$ standing for average density and diameter respectively, subscripts
$S$ and $m$ for Sun and Moon, we have
\beq\label{XX}
{\rho_S}{\left(D_S\over r_S\right)^3}=
{f \rho_m}{\left(D_m\over r_m\right)^3}.
\eeq
But $D/r$ is just the apparent angular size $\theta$ of the object subtended at the Earth.
This means, for example,
that the total tidal potential at latitude $l$ and time $t$
may be written
\beq
\Phi =GR_E^2[ \rho_S\theta_S^3 A(l,t) + \rho_m\theta_m^3 B(l,t)]
\eeq
where $G$ is the gravitational constant, $R_E$ the radius of the Earth, and
$A$ and $B$ angular functions of order unity ($t$ serves as a longitudinal variable).
The relative Sun and Moon contributions are thus extremely sensitive to their
relative angular sizes.   Equation (\ref{XX}) implies
\beq
\theta_S = \left(f\rho_m\over \rho_S\right)^{1/3} \theta_m.
\eeq
Were the densities the same, equal angular sizes would mean equal tides.
With $(\rho_m/\rho_S)^{1/3}= 1.34$, we see that values of $f$ even roughly near 0.5
will translate
to nearly equal angular sizes for the Sun and Moon subtended at the Earth.
The question of why these angular sizes should be so closely matched now becomes one of why
the Sun's tides should be something like half of the Moon's.   The ground has shifted from
perception to dynamics, and dynamics has calculable consequences.

\section{Analysis}
\subsection{Semi-qualitative considerations}

There are many characteristic frequencies in
the tidal problem, the precise number depending upon the level of
accuracy and time baseline sought.  The three most important
frequencies are associated with the diurnal rotation period of the
Earth, the siderial orbital period of the Moon, and the yearly
passage of the Sun along the ecliptic.  We
experience events on the rotating surface of the Earth, which
boosts the effective forcing frequencies of the Moon and Sun by $2\pi/(1\
{\rm day})$.  For the problem at hand this is a large number, and the
difference between the effective lunar and solar forcing frequencies is
small compared to their average.  When two processes
with close but unequal frequencies superpose, the net response is
carried at this average frequency, with a modulation envelope at the
difference frequency.  In the case of the tides the modulation is
particularly rich, because there are many different modulation
frequencies that enter.  By contrast, were one of the solar or lunar
tides heavily dominant, very little of this modulation
richness would be present.  It is this feature of the problem that
compels one to consider the importance of comparable tidal contributions
from the Sun and Moon.

The close match of the Sun and Moon angular size combined with the density
ratio implies that the solar tidal contribution is somewhat smaller
than the lunar.  Amplitude modulation of the tidal forcing would
still be present if the inequality of magnitudes were reversed; might
it just as easily have occurred that the Moon's contribution was
smaller than that of the Sun?  At an orbital radius of 1 AU
the net tide would then of course be considerably smaller, 
and if the Earth is in fact
near the inner edge of the the Sun's habitable zone as some
current estimates suggest, a more distant orbital location would
make the tides yet smaller.   
If a putative moon with a mass comparable to the true Moon had to form
close to the Earth by impact or otherwise, there is a nautural
link between relative and absolute tidal amplitudes.  
The early moon's tidal
contribution, putatitive or real, would have been overwhelming,
moving the Earth's crust through kilometer-scale upheavals.  The
lunar orbit would evolve rapidly at first, with the satellite spiraling
outward via tidal dissipation to several $10^5$ km, where the orbital
recession would slow to a comparative crawl, conveniently measured
in centimetres per year.  In the process, not only would the tides
greatly diminish to below environmentally friendly meter-scale
displacements, {\it at the same epoch} the Sun would also become a player
in the tidal game.  In other words, for any Moon-like
satellite, orbital dynamics lead to
something resembling the present tidal environment: the Moon
dominant but not overwhelmingly so.

The scenario might be different if a moon formed farther out and evolved
{\em inward}, but this is not possible if the host planet rotates more rapidly than
the moon orbits, as presumably would be the case on a planet capable of
supporting life.  If the planet rotates more rapidly, the effect of tidal dissipation
on the perturber's orbit is to cause outward migration.  (It is generally
thought the absence of
of any moons associated with the two slowly rotating inner planets Venus and Mercury
is due to the resultant {\it inward} tidal migration of any primordial moons
circling these bodies [4].)

\subsection {Tidal Potential}
\subsubsection{Coordinate expansion}

To calculate the actual height of the oceanic tides at a particular location is a
difficult task.  The answer depends on the details of the coastline in question
(especially whether resonances are present),
the depth of ocean as a function of position (bathymetry),
the hydrodynamics of ocean-propagating long waves (shallow water waves modified
by the Earth's rotation), and coefficients known as the Love numbers,
used to extract the difference between the oceanic and solid crust
tidal response.   
We surely do not know Devonian bathymetery or the the Laurussian coastline
in sufficient detail to perform a precise calculation of this type.

Fortunately, extreme accuracy is not required.
A quantity known as the {\em equilibrium tide} will suffice.
This is simply the displacement of
a local equipotential surface caused by the introduction of the tidal potential,
and is directly proportional to this potential.
The displacement is calculated by setting the associated work done against the Earth's gravity
equal to (minus) the disturbing tidal potential.

Let $\Phi_t$ be this potential and $g$ be Earth's gravitational field.
The height $h$ of the tide is then simply
\beq
h = -{\Phi_t\over g}.
\eeq
The equilibrium tide is a reasonable measure of the scale of the reponse.  In any case,
we are less interested in $h$ as an actual height
then the fact that it is proportional to $\Phi_t$,
the driving potential of the problem.  We are particularly interested in
the temporal behaviour of $\Phi_t$, and in effect use $h$ as a convenient normalization.

The calculation of $\Phi_t$ is a standard exercise [4] and not difficult.  We
briefly review it here to keep the presentation self-contained.
Let the centre-of-mass distance between the Earth and Moon be $r_m$.  At the centre of the
Earth we erect a $\bb{u}=(u,v,w)$ coordinate system, where $w$ is the distance along the
line connecting the centres-of-mass, and $u$ and $v$ are orthogonal axes oriented
so that $u,v,w$ is a standard right-handed Cartesian system.   The potential at
the coordinate location $u,v,w$ is
\beq
\Phi = - {GM_m\over [(r_m+w)^2 +u^2+ v^2]^{1/2}}
\eeq
It is important to note that while the origin of our $uvw$ system is fixed to the Earth's
centre, the axes are not fixed to the Earth.  They
are defined by the Moon's instantaneous location (see below.)

Since $r_m$ is large compared with $u$, $v$, or $w$, we expand $\Phi$ through
second order in small quantities:
\beq\label{t1}
\Phi = {GM_m \over r_m}\left[ -1 +{w\over r_m}+{1\over 2r_m^2} (u^2+v^2-2w^2)\right]
\eeq
The first term is an additive constant contributing
nothing to the force $-\del\Phi$.  The gradient of the
second term returns the dominant $1/r^2$ force along
the ($w$ axis) centres-of-mass, and the third term is the desired leading order tidal potential $\Phi_t$, giving
rise to a force vector proportional to $(-u, -v, 2w)$.  Relative to the Earth's centre,
the $w$ force is repulsive and the $uv$ forces are attractive.
We thus find
\beq\label{YYY}
h_m = {GM_m\over 2gr_m^3} \left( 2w^2-u^2-v^2\right)
\eeq

The more cumbersome calculation arises when one relates
these $uvw$ coordinates, defined by and moving with the Moon's orbit, to
coordinate axes {\em fixed} to the rotating Earth.  Let us refer to these Earth body axes
as $\bb{x_b}=(x_b, y_b, z_b)$, with their origin at the Earth's centre and
the $x_b$-axis parallel to the North Celestial Pole.
With $\phi_m(t)$ equal to the azimuthal angle of the Moon
in its own orbital plane, $i_m$ the Moon's orbital inclination relative to the
equator, $\alpha$ the shift in the azimuth of the line of nodes (i.e., the line formed
by the intersection of the equatorial and lunar orbital planes),
$\Omega_E$ the Earth's diurnal angular rotation rate, and $T$ denoting transpose
(so the vectors are in column form) the transformation is given by
\beq\label{YY}
\bb{u^T} = \bb{\cal R_X}(\phi_m)\bb{\cal R_Y}(i_m)\bb{\cal R_X}(\alpha)\bb{\cal R_X}(-\Omega_Et)\bb{x^T_b}
\eeq
where $\bb{\cal R_X}(\theta)$ is a $3\times 3$ rotation matrix about the $x$ axis
\beq
\bb {\cal R_X}(\theta) =
 \begin{pmatrix} 1&0&0\\
0&\cos\theta&\sin\theta   \\
0&-\sin\theta   &\cos\theta  
\end{pmatrix}
\eeq
and $\bb{\cal R_Y}(\theta)$ is a $3\times 3$ rotation matrix about the $y$ axis
\beq
\bb{ \cal R_Y}(\theta) =
 \begin{pmatrix} \cos\theta&0&\sin\theta\\0&1&0\\
-\sin \theta &0&\cos \theta\end{pmatrix}
\eeq
(Note that $i_m$, $\alpha$ and $\phi_m$ are in effect Euler angles.)
By
substituting (\ref{YY}) into (\ref{YYY}), we may determine the potential at a particular
terrestrial location.   Exactly analogous formulae hold for the solar orbit.

\subsubsection {Eccentricity}

A complication of practical importance is the eccentricity of the Moon's orbit.
Its present value is [5]:
\beq
\epsilon_m = 0.0549.
\eeq
The $1/r_m^3$ behaviour of the tidal amplitude means that there is modulation
from this effect alone.  A more minor modification is that the temporal advancement of the
azimuth in the orbital plane is no longer uniform.  With $\Omega_m$ equal to the
average orbital frequency, the first order correction is:
\beq
\phi_m(t) = \Omega_m t - \varpi + 2\epsilon_m\sin(\Omega_m t- \varpi)
\eeq
where $t$ is time and $\varpi$ is the longitude of the pericentre.
The current solar eccentricity is $\epsilon_s=0.0167$ [5].
The separation $r_m$ is given by the usual formula [23]:
\beq
r_m = {r_0\over 1 + \epsilon\cos\phi_m(t)}
\eeq
where $r_0$ is the semilatus rectum of the orbit. Analogous formulae hold for the Sun.

\begin{figure}
\includegraphics[width=16cm]{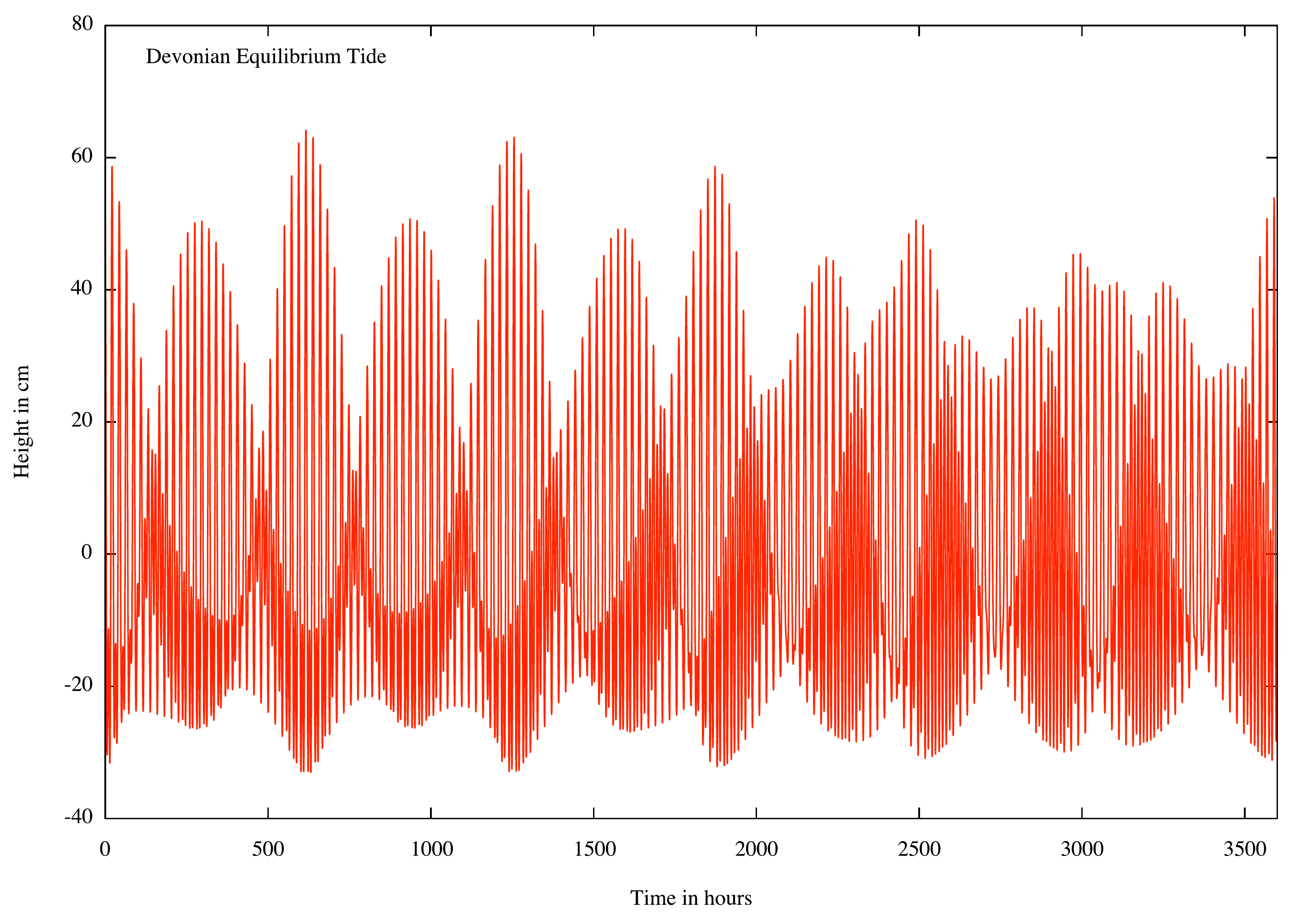}
\caption {Representative Devonian equilibrium tide ($h$ in cm) at latitude $35^\circ$ S
for a lunar semi-major axis of $365000$km,
a plausible value for the late Devonian Period.  Other orbital parameters correspond to current
values, except for arbitrarily chosen longitudes of the pericentres.  (See text.)
The displayed baseline is 150 days. Note that the driving is indeed highly modulated.}
\end{figure}
\begin{figure}
\includegraphics[width=16 cm]{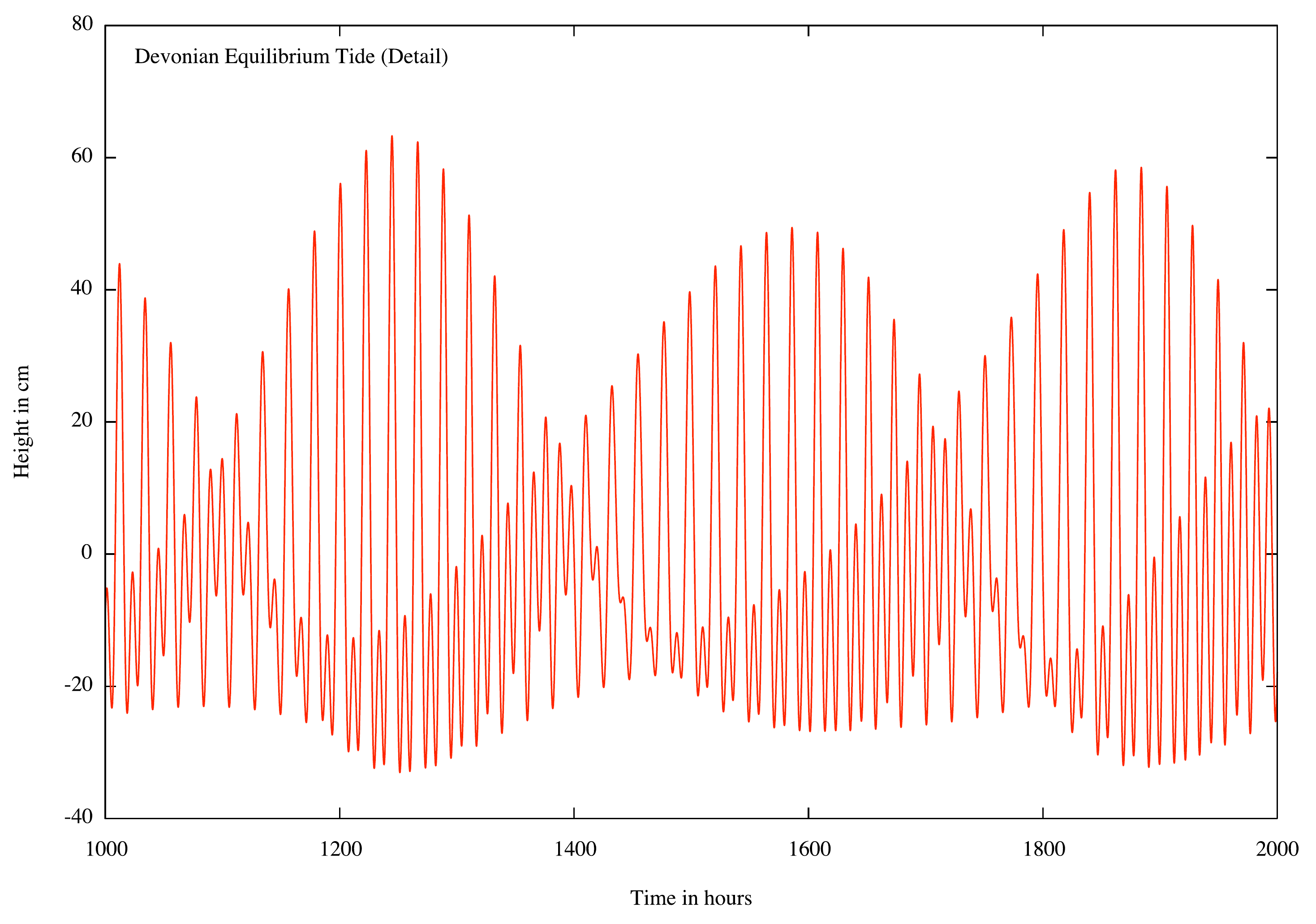}
\caption{Detail of previous figure between 1000 and 2000 hours.}
\end{figure}

\subsection {Equilibrium tide}
In figure (1), we show the total equilibrium tide, $h(\bb{x_b})= h_m(\bb{x_b})+h_S(\bb{x_b})$.
For this canonical case, we
have chosen parameters for the solar orbit to be those of today.   For the lunar orbit
we use an estimated Devonian semilatus rectum 
of 365000 km [4,6].
(The Earth's rotational frequency has
accordingly been increased
to conserve total angular momentum.)   We have used the current orbital inclinations and
eccentricities.   It is certainly possible that these have evolved for the Moon [7],
and the Milankovitch cycles affect the Earth's orbit about the Sun,
but our results
are not sensitive to nonpathological values.   The lunar and solar longitudes of the
pericentres have been chosen arbitrarily ($0.63\pi$ and $-0.17\pi$ respectively); the precise values
are inconsequential.  The $\alpha$ angle currently rotates with an 18.6 year
period.   Its precise value is also not critical, though small values show particularly sharp
modulation.  The canonical value is thus $\alpha=0$.
The latitude chosen, $35^\circ$ S, is supposed to be representative
of the late Devonian Laurussian coast.   The most striking feature
of $h$ is the presence of many different incommensurate
frequencies, the most important of which are diurnal, semi-diurnal and
twice the lunar orbital frequency.  Note the asymmetry between
the diurnal and semi-diurnal components, which results in a 
``shading'' effect in the lower
portion of the figure.  (The  asymmetry is caused by the orbital inclinations.)
A higher time resolution detail is shown in figure (2) between 1000 and 2000 hours.
The strength of the diurnal equilibrium high tide can
vary by a factor of 5:  it is highly modulated.

There is a self-evident
repeating pattern of a build-up of the tidal strength to a maximum,
corresponding to the deepest inland penetration of the sea line,
followed by a recession, which can be sharp, in which the inland penetration diminishes each day.
It is not the height of the spring tide (say, at 600 hours) that is relevant here.  Rather, it is the
rapid rate of change of the ``modulation envelope'' (at about 700 hours).    Being a maximum, the spring tide changes
little from one local peak to the next, but the changes of the high water mark just after the
spring tide become treacherous.     
The waxing and waning shoreline
would have left behind a series of tidal pools of increasing depth
(or at least more frequent replenishment)
with decreasing distance from the sea.  Any aquatic tetrapod stranded
in a shallow inland pool that was
adapted to squirming to a nearby reservoir
would clearly have been favoured over those less mobile.   This
selection pressure was likely to have presented itself relentlessly.

\begin{figure}
\includegraphics[width=16cm]{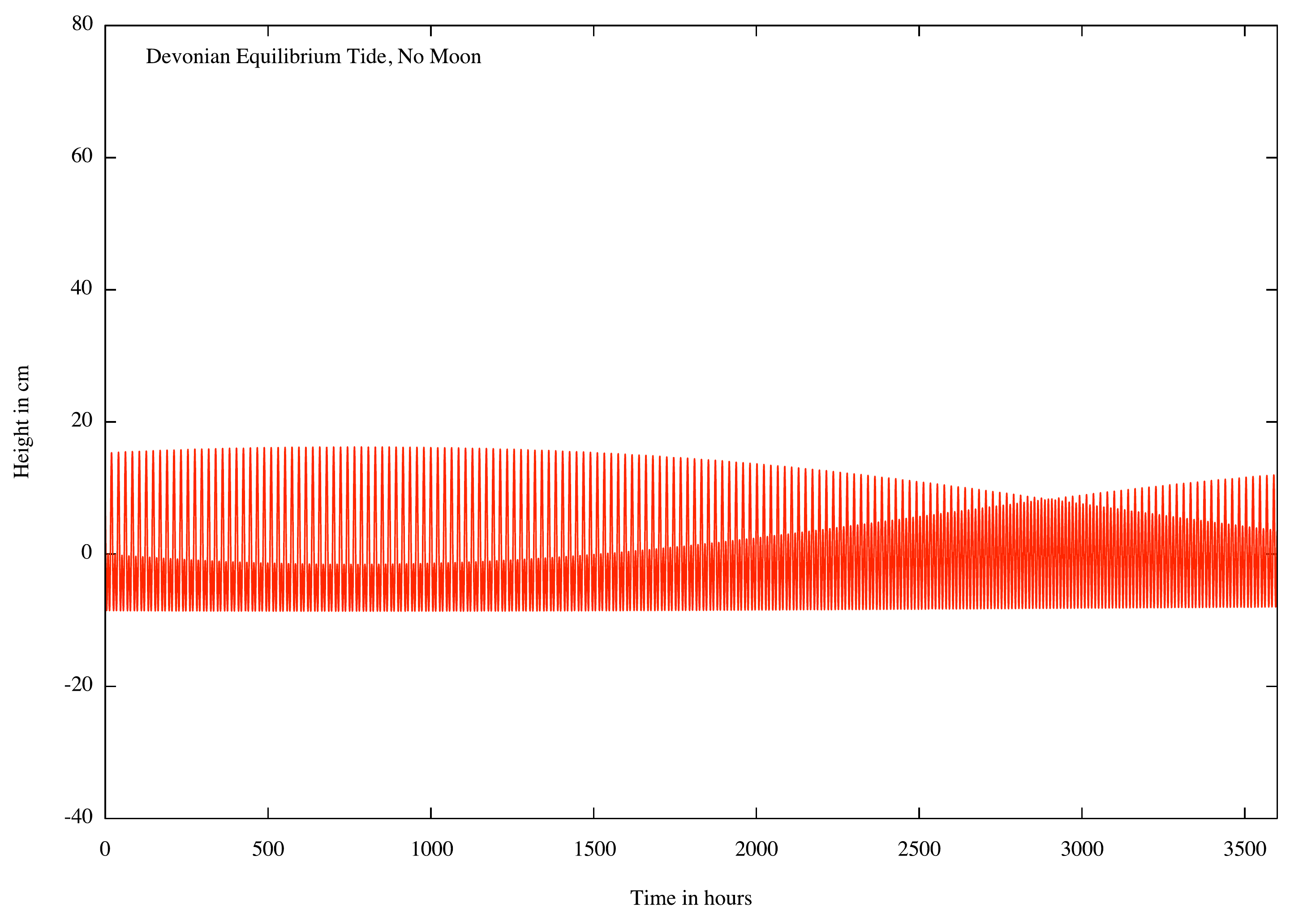}
\caption {Equilibrium tide without the Moon's contribution.   The contrast
with figure [1] is evident.}
\end{figure}

This
should be contrasted with the equilibrium tide that would
obtain if there were no (or a tidally unimportant) Moon (figure [3]).
This tide would be one-third of the amplitude and much less variable.
Local topography (and weather!) might still lead to the stranding of aquatic life
forms of course, but probably not with an extensive network of pools
leading back to the sea.  Whether this would lead to a different
course of evolution is a matter of speculation, but there is a real
qualitative difference in tidal ponds and estuarial flooding
in these two scenarios.

Finally, in figure (4) we show for comparision the equilibrium tide for the cases corresponding
to a lunar angular diameter half that of the Sun's, without
changing the Moon's average density.   There is some modulation, but far more gentle than
and qualitatively different from our canonical case.   It is only when the angular sizes
become close that we start to see highly sculpted modulation.

\begin{figure}
\includegraphics[width=16cm]{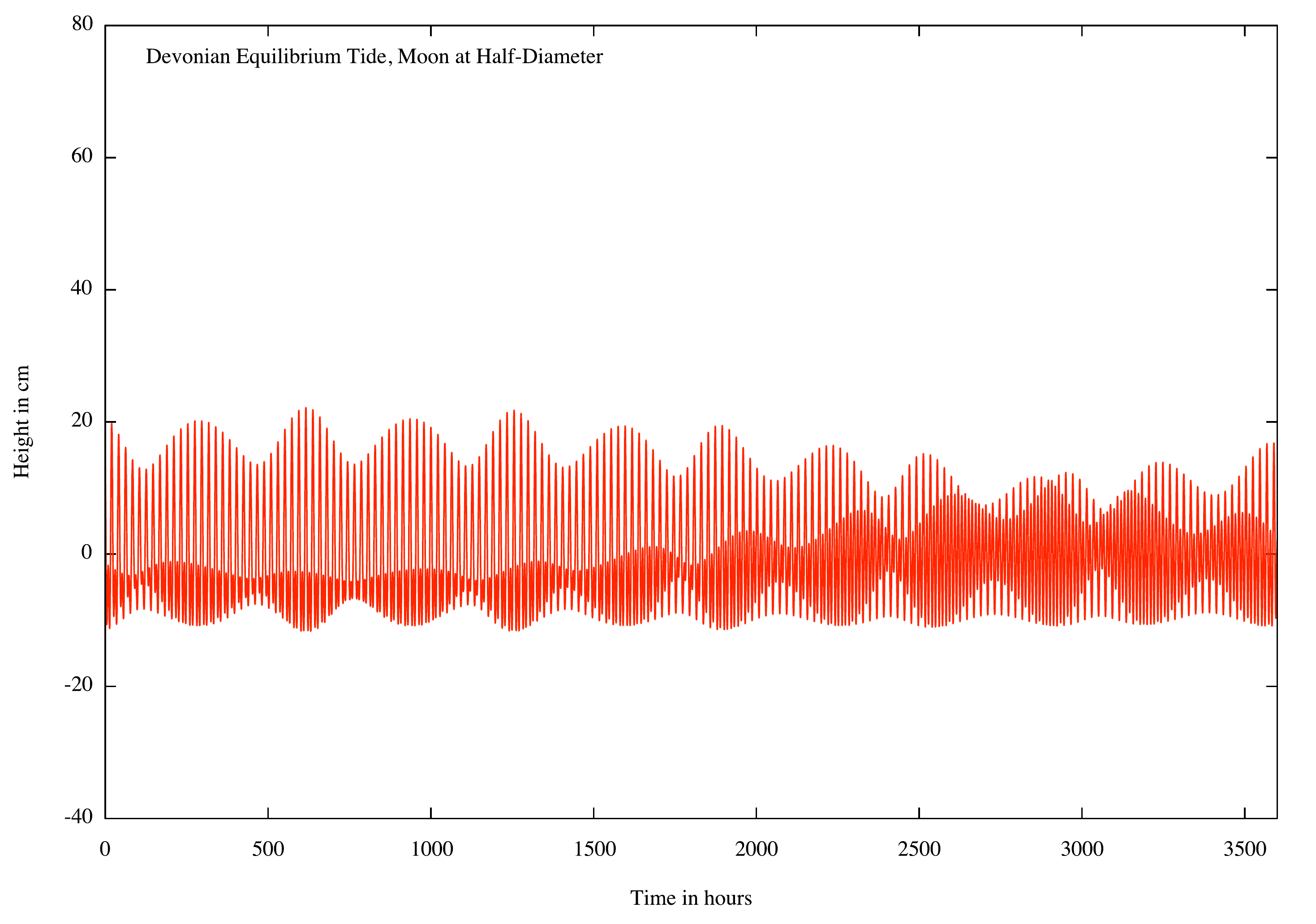}
\caption {Equilibrium tide for the case of the Moon subtending half the angular diameter
of the Sun, assuming an unchanged average lunar density.  There is only very gentle amplitude
modulation.}
\end{figure}

\section {Discussion}
\subsection {Tidal receptivity}

Any discussion of 
continental reconstruction prior to 200 Ma must begin by noting that this 
is an unavoidably speculative undertaking.  Nevertheless, there seems
to be some consensus on major features.

At the time of the middle Silurian (430 Ma), the broad intercontinental
seaway comprising the Rheic Ocean separated the (southern) Gondwanan
and (northern) Laurussian land masses.  The early Devonian marked the
beginning of the
closure of this seaway, a protracted geological event.
The squeezing of the Rheic was not uniform
along its length, however.  Some reconstructions show
the eastern Rheic squeezed down to brackish
swampland, while in the west
the Rheic maintained a very broad opening of several thousand kilometers
before giving on to the great Panthalassa Ocean.  More recent
reconstructions suggest a western closure preceeding the eastern [8].
In either scenario, a tapered,
horn-shaped configuration of the intercontinental seaway was
maintained throughout most of the Devonian, before the Rheic
became closed off at the start of the Carboniferous, eventually
uniting the Laurussian and Gondwanan land masses, forming the Pangaean
supercontinent of the Permian.

From the point of view of tidal dynamics, the Devonian is distinctive.
The intercontinental seaway dividing Laurussia from Gondwana had the same
generic form as the current Bay of Fundy, Bristol Channel, or northwest
coast of Australia:  a broad opening onto a deep ocean, tapering
to shallower seas.   These are all regions known for
their large tidal range.  The propagation speed of water waves
is greater in deep water than in shallow seas, going roughly as the
square root of the water's depth [9].   In the case of interest, this means
that the shallows would not have responded as rapidly to tidal forcing as would the
deeper water in the opening to the
ocean.  The resulting tidal surge propagates into the shallows,   
the flow convergent because of the tapering of the channel.
The consequence of this behaviour is a greatly enhanced rise of the propagating tide.
A correspondingly rapid egress occurs at low tide.
This behaviour accounts for the famously large tidal ranges of the modern regions noted above.

The Devonian, we therefore expect, was likely to be have seen dramatic
tides along the Rheic Ocean, with both significant modulation of the high tide
level and a great tidal range.   It is not difficut to imagine the resulting
rich network of coastal tidal pools that is likely to have been present.  It is therefore
highly suggestive that the earliest identified tetrapod trackways are thought to have
originated in the Eifelian stage of the Devonian, significantly predating body
fossil tetrapod remains, at a paleogeographic location
corresponding to a tight constriction of the central Rheic seaway [10].
The southern tropical coasts of the Devonian Earth may well have been a massive swamp.

\subsection{A brief overview of Devonian tetrapodomorph habitats}

One of the richest sources of Devonian tetrapodomorph fossils is a
huge area extending through parts of modern day Lithuiania, Latvia, Estonia
and Russia.  In Devonian times, it was a massive delta region, with 
clear evidence of tidal influence in the form of currents and facies [11].  
Yet more telling, from the point of view of this work, is the evidence of
interruption of river current sediment deposition by tidal currents. 
The delta plane was graded.  In the upper plane region,
the evidence indicates sporadic interruption during spring tides;
the lower plane region shows a much more regular pattern of tidal currents.
(This is analgous to the modern day River Severn, which hosts tidal bores at 
spring tides [9].)   The Baltic Devonian Delta thus preserves explicit evidence for
the environmental influences of {\em modulated} tidal forcing.  The formation
existed for some 30 million years, and for much of this time provided habitats
for tetrapods and near-tetrapods [11, 24].

More generally, the distribution of Devonian tetrapodomorph fossils 
indicates a preference for marginal marine environments [2]; the {\em
extent} of the distribution implies an ability to cross narrow
seaways.  It is of interest to consider also
elpistostegids, such as {\em Pandrichthys,} {\em Livoniana} and {\em Tiktaalik,} the
fish group from which the tetrapods are thought to have evolved.
While
they retained paired fins (not limbs), they probably enjoyed some
terrestrial maneuverability.  Elpistostegid fossils extend
further back in time than the earliest true tetrapod body fossils,
but do not predate the earliest known tetrapod trackways [10].  Evidently,
there was an extended period of coexistence between the two groups.
From the mid-Devonian, fossils of the more primitive tetrapodomorph
{\em Eusthenopteron} have been found in coastal marine sediments [2, 13]
(Eastern Canada), as have {\em Panderichthys} and
{\em Livoniana} fossils (Latvia) [12]; 
{\em Tiktaalik} remains are found in non-marine fluvial deposits (Ellesmere Island, Canada)
from the same period [14].   The former environment would
very likely have been subject to tidal influences; the latter is less
certain but by no means impossible.
The case for a tidally-influenced environment
also applies to the tetrapods {\em Elginopteron} [15] (non-marine fluvial
deposits, Scotland) and {\em Tulerpeton} [16] (coastal lagoon, Russia).  By contrast, 
{\em Ichthyostega} [17] and {\em Acanthostega} [18], the earliest known tetrapods
with fairly complete body fossils (later than {\em
Elginopteron}, for which only fragmentary remains are known) are
associated with a non-marine inland basin (Greenland), a more ambiguous tidal
zone.  Remains of the approximately contemporaneous tetrapod {\em Ventastega} [19]
were found in river bed tidal deposits (Latvia).   Note that
{\em Ventastega,} {\em Panderichthys,} {\em Livoniana} all left
remains in the massive Baltic Devonian Delta system described above.
Finally, the Eifelian
trackway is an important datum, as it represents the earliest known
evidence of tetrapod activity.  The trackway was found in sediments
associated with a coastal lagoon (Poland) [10].

The fossil record generally supports the notion that tidal
modulations contributed to the shaping of the environment of tetrapodomorphs.  

\section{Conclusion}

The very near angular sizes of the Moon and Sun as seen from the
Earth are a mathematical by-product of the existence half-meter,
highly modulated, quasi-periodic equilibrium tides associated with
a planet of order 1 AU from
a Sun-like star.  These conditions
have been examined quantitatively in this work
by explicit calculation of the equilibrium tides under a variety of different
assumptions.  It is probably rare for a planet to harbour highly
complex macroscopic organisms (though hard data on this are of course scarce!),
and it must also be unlikely for a planet to have
a large moon nearly matching the central star in angular diameter.
If these outlandish features are unrelated, why should the same planet just
happen to have both?  What is certain is that the Sun and Moon both are able to
contribute significantly to the net tide, that this introduces very strong amplitude
modulation effects that would otherwise be absent, and that early
tetrapods would have had to cope with becoming stranded in a constantly changing network
of shallow inland tidal pools.  
The uncertainty is whether these ineluctable
consequences of strong tidal modulation were 
essential, or merely incidental,
to creating an evolutionary pathway leading to a contemplative species.

It may be just a coincidence that our planet has all these
features in common for no particular reason.  But this line
of argument doesn't sit well, and is in any case totally sterile.
In terms of the sheer number of phyla and diversity of species, the Earth's intertidal
zones are among the richest habitats on the planet [2].
Despite their ostensible stranding hazards, these regions stimulated diversity, not avoidance.
It seems reasonable to
consider the notion that not just the existence of the tide, but its
particular form, may have influenced the
course of evolution, selecting for (among other things),
efficient maneuverability and motility in networks of
shallow tidal pools.  A.S. Romer's classic vision of trapped tetrapods
striving for accommodating pools is supported by the apparent coincidence
in angular sizes of the Sun and Moon.  The fact
that the resulting tidal pools would not have been in arid zones---one of the early criticisms
of part of Romer's theory that has allowed it to fall into disfavour---is
irrelevant, if isolated shallow pools are common because of the tidal
dynamics.  As has been noted elsewhere [20], aridity is really not an issue.
Puddles can drain or be rendered unsuitable for habitation under humid conditions as well.
For Romer's purposes, what is really needed is a developed network of ponds, and this is what 
the dynamics of modulated 
tides provides.   A striking contemporary example of pond-searching is
evinced by the so-called ``climbing perches'', {\it Anabas testudineus} [21], air
breathing fish who
literally save themselves by terrestrial locomotion from one drying puddle to 
a deeper pool.  These fish, whose behaviour constitutes a sort of Romerian ideal,
inhabit wetlands in southeast Asia, hardly an arid climate.  It is not a
great conceptual 
leap to envision a similar survival imperative (with no contemporaneous land-based
predators) in Devonian swamps.     

There may also be interesting geophysical
consequences of noncommensurate modulated tides acting over billions
of years that have yet to be explored---the Earth's Love numbers, by which
one measures the solid planetary response are by no means tiny.
Yet further afield, it is of interest to note that
the search for the moons of extrasolar planets (``exomoons'') is now in its
infancy, and is expected to return significant results in the next few years [22].
The discussion of this paper suggests a special role for those
moons providing a tidal force comparable to the planet's host star.
For if it is necessary to have the sort of heavily modulated tides we experience on
the Earth in order to influence a planet's evolutionary course in a manner
constructive for evolving complex land-based organisms,
the mystery of nearly equal
angular sizes of the Sun and Moon would evaporate, rather like an inland
Devonian tidal pool.

\section*{Acknowledgements}

The author has benefitted greatly from interactions with many colleagues over
the course of this work.
He is deeply grateful to A. Lister for his willingness to guide
an outsider through pertinent biological literature and for critical advice,
and to P. Ahlberg for an exceptionally meticulous review and extended correspondence.  
He would also like to thank D. Lynden-Bell for drawing the author's attention to
Newton's early work on tides; C. Sarazin for sending his unpublished manuscript;
D. Balbus, R. Harvey and M. Rees for stimulating conversations;
and J. Clack and R. Dawkins for their helpful advice and active encouragement.
Support from the Royal Society in the form of a Wolfson Research Merit Award
is gratefully acknowledged.



\begin{thebibliography}{99}

\bibitem {1} Romer, A. S., 1933, Man and the Vertebrates, (University of Chicago Press: Chicago).
\bibitem {2} Clack, J. A., 2012, Gaining Ground:
The Origin and Early Evolution of Tetrapods, (Bloomington: Indiana Univ. Press).

\bibitem{3}
Tolbert, C. S., \& Sarazin, C. L. solar eclipses, tides,
and the evolution of life on the Earth, 1992, unpublished manuscript.

\bibitem{4}
Stacey, F. D., \& Davis, P. M., 2008, {Physics of the Earth,} (Cambridge University
Press: Cambridge).

\bibitem{5}  Allen, C. W., 2000,
{Astrophysical Quantities,} (Springer-Verlag: New York) (2000).

\bibitem{6}  Williams, G. E., 2000,
Geological Constraints on the Precambrian History of the Earth's Rotation and
the Moon's Orbit, {Rev. Geophys.} 38, 37-59.


\bibitem{7}
Garrick-Bethell, I.,
Wisdom, J., \& Zuber, M.\ T., 2006, Evidence for a Past High Eccentricity Lunar Orbit,
Science, 313, 652-655.

\bibitem {8} Golonka, J. \& Gaweda, A., 2012, Plate Tectonic Evolution of the Southern
Margin of Laurussia in the Paleozoic, in Tectonics--Recent Advances, E. Sharkov ed.,
Ch. 10, pp. 261--282. (InTech: ISBN 978-953-51-0675-3, doi 10.5772/2620)

\bibitem{9} 
Lighthill, J., 1978
{Waves in Fluids,}(Cambridge University Press: Cambridge), p. 442.

\bibitem{10} Nied\'zwiedzki, G.,
Narkiewicz, K., Narkiewicz, M., \& Ahlberg, P. E., 2010, Tetrapod trackways from the early
Middle Devonian period of Poland, Nature, 463, 43-48, (doi10.1038 nature08623).

\bibitem {11}
Pont\'en, A., \& Plink-Bj\"orklund, P. 2007, Depositional environments in an extensive tide-influenced
delta plain, Middle Devonian Gauja Formation, Devonian Baltic Basin, Sedimentology, 54, 969--1006.

\bibitem{12} Ahlberg, P., Luksevics, E., Mark-Kurik, E. 2000, A near-tetrapod from the Baltic 
Baltic Middle Devonian, Palaentology, 43(3), 533--548.

\bibitem {13}
Schultze, H.-P., \& Cloutier, R. (eds.) 1996, Devonian fishes and plants of Miguasha, 
Quebec, Canada, (Friedrich Pfeil: Munich).  

\bibitem{14}
Daeschler, E. B., Shubin, N. H., Jenkins, F. A. 2006, A Devonian tetrapod-like fish and the evolution
of the tetrapod body plan, Nature, 440, 757--763.

\bibitem{15}
Ahlberg, P. E. 1991, Tetrapod or near tetrapod fossils from the Upper Devonian of
Scotland, Nature, 354, 298--301.

\bibitem{16}
Alekseyev, A. A., Lebedev, O. A., Barskov, I. S., \& Kononova, L. I., \& Chizhova, V. A. 
1994, On the stratigraphic position of the Famennian and Tournaisian fossil vertebrates in 
Andreyevka, Tula Region, Central Russia, Proceedings of the Geologists Association, 105, 41--52.

\bibitem{17}
Jarvik, E. 1952, On the fish-like tail in the ichthyostegid stegocephalians, Meddelelser om
Gr{\o}nland, 114, 1--90.


\bibitem{18} 
S\"ave-S\"oderbergh, G. 1932, Preliminary note on the Devonian stegocephalians from East 
Greenland, Meddelelser om Gr{\o}nland, 98, 1-211.

\bibitem {19}
Ahlberg, P., Luksevics, E., \& Lebedev, O. 1994, The first tetrapod finds from the Devonian
(Upper Famennian) of Latvia, Philosophical Transactions of the Royal Society of London B, 
343, 303--328.

\bibitem{20}
Dawkins, R. 2004, The Ancestors Tale, (Houghton Mifflin Company: New York), p. 345.

\bibitem{21}
Hughes, G.\ M. \& Singh, B.\ N., 1970, ``Respiration in an Air-Breathing Fish,
the Climbing Perch {\it Anabas testudineus},'' Journal of Experimental Biology, 
53, 265-280.



\bibitem{22}
Kipping, D. M., Bakos, G. \'A.,
Buchhave, L., Nesvorn\'y, D., \& Schmitt, A., 2012,
The Hunt for Exomoons with Kepler (HEK). I. Description
of a New Observational Project, {Astrophys. J.} {750:115} (19pp).



\bibitem{23}
Murray, C.\ D., \& Dermott, S.\ F., 1999, Solar System Dynamics, (Cambridge University Press:
Cambridge), p. 28.



\bibitem{24}
Ahlberg, P., Clack, J. A., Luksevics, E., Blom, H., Zupins, I. 2008, {\em Ventastega curonica} and the
origin of tetrapod morphology, Nature, 453, 1199--1204.


\end{thebibliography}
\end{document}